\documentstyle[12pt,here,epsf,rotate]{article}
\newcommand{\bel}[1]{\begin{equation}\label{#1}}
\newcommand{\ee}{\end{equation}}

\def\oscsigmaqq{{\textstyle \sum }_{qq} ^{\rm osc} }
\def\oscsigmapp{{\textstyle \sum }_{pp} ^{\rm osc} }
\def\dpr{{\prime \prime}}
\def\dtwohqzero{\left<{\partial^2\hat H\over \partial Q^2}(Q_0)
    \right>^{\rm qs}_{Q_0,T_0}} 
\def\corrheat{_0\psi^{\dpr }}
\def\self{{\it\Sigma}}
\def\selfre{\Sigma^\prime}
\def\Qtzer{Q(t_0)}

\def\Ptzer{P(t_0)}
\def\fext{f_{\rm ext}}
\def\qext{q_{\rm ext}}
\def\respc{\chi_{\rm coll}(\om)}
\def\respposc{ \chi _{\rm osc}^{\prime \prime } (\omega )}
\def\men{{\cal E}}
\def\delmen{\Delta {\cal E}}
\def\staw{{\it \Omega}}

\def\hbo{\hbar \om}
\def\fmb{ \langle \hat F \rangle }
\def\fav#1{ \langle \hat F \rangle _{#1}} 
\def\h#1{{\hat #1}}
\def\ham#1{\hat H(\xvi,\pvi,#1)}
\def\ffield#1{ \hat F(\xvi,\pvi,#1)}
\def\h#1{{\hat #1}}
\def\Vres{\hat V^{(2)}_{\rm res}(\xvi ,\pvi )}
 
\def\equiop{\hat{\rho}_{\rm qs}}

\def\rfc#1{\chi_{{\rm coll}} (#1)}
\def\drf{\chi^{\prime \prime }}

\def\drft{{\widetilde{\chi}}^{\prime \prime }}

\def\rft{{\widetilde{\chi}}}
\def\resqq{ \chi _{qq} (\omega ) }
\def\chiosc{\chi_{\rm osc}(\om)}

\def\totfren{{\cal F}}
\def\bra{\big\langle}
\def\ket{\big\rangle}

\def\sigmaqqini{ {\textstyle \sum }_{qq} \! (t_0) }
\def\sigmaqpini{ {\textstyle \sum }_{qp} \! (t_0) }

\def\sigmappini{ {\textstyle \sum }_{pp} \! (t_0) }
\def\sigmaqq{ {\textstyle \sum }_{qq} \! (t) }
\def\sigmaqp{ {\textstyle \sum }_{qp} \! (t) }

\def\sigmapp{ {\textstyle \sum }_{pp} \! (t) }

\def\eqsigmaqp{{\textstyle \sum }^{eq}_{qp} }
\def\eqsigmaqq{{\textstyle \sum }^{eq}_{qq} }
\def\eqsigmapp{{\textstyle \sum }^{eq}_{pp} }
\def\qct{Q^c(t)}
\def\pct{P^c(t)}
\def\mev{\;{\rm MeV}}
\def\xvi{{\hat x}_i}
\def\pvi{{\hat p}_i}
\def\om{\omega }

\author{H.~Hofmann 
\thanks{e-mail: hhofmann@.physik.tu-muenchen.de}
\thanks{http://www.physik.tu-muenchen.de/tumphy/d/T36/hofmann.html}\\
        Physik-Department, TU M\"unchen, D-85747 Garching
        \vspace{0.8cm}}
\title{Nuclear Transport at Small Excitations:
Thermal and Quantal Aspects
\thanks{Prepared for the RIKEN Symposium on "Dynamics in Hot
Nuclei", Tokyo, March 1998}
  \vspace{0.5cm}}

\begin{document}
\maketitle

\begin{abstract}
\noindent
The application of the locally harmonic approximation to large scale
collective motion is briefly reviewed. Particular emphasis is paid to
issues which might be useful in the more general context, or which are
specific to our treatment, like there are: Self-consistency between
collective and intrinsic motion, a $T$ dependent coupling constant, the
inclusion of quantum effects for the dynamics of fluctuations. Finally,
open problems are addressed, like the use of thermodynamic concepts at
smaller excitations, the ergodicity question, the variation of
transport coefficients with the nuclear shape due to shell structure,
the limits of the LHA at smaller $T$. 

\end{abstract}

\section{Introduction}

In this talk I would like to briefly report on the transport
theory which has been worked out on the basis of a suitable
application of linear response theory. In the second part I
shall address some of the most stringent and largely unresolved
problems one generally faces in dealing with nuclear transport at
smaller excitations. Details will be presented only to outline
the general concepts. For more extensive studies I like to
refer to my recent review article \cite{hofrep} as well as to the
original papers, especially to those of the more recent years
which have been published together with F.A. Ivanyuk, D.
Kiderlen and S. Yamaji and others.

\subsection{The basic approximation scheme}
Unfortunately, still too often theories of nuclear collective
motion are considered separately from transport models. This is
unjust both in the light of present day experiments as well as
from the more general point of view. Indeed, it is probably only
for rotational motion exactly along the Yrast line that
intrinsic excitations of the nuclear system do not play any
role. Moreover, one should not forget that one of the first
theoretical descriptions of the prime example of large scale
collective motion, nuclear fission, was given by Kramers
\cite{kram} as early as 1940 and which was based on the picture of
"transport" in collective phase space. In his famous equation
there appear such concepts as dissipation and stochasticity.
(The reader who is not familiar with this equation is asked to
wait for a short while; we will come back to it later).  Kramers
applied this equation to the decay rate for fission, to find a
generalization of the famous Bohr-Wheeler formula. In these days
the origin of dissipation was attributed to the strong
"correlations" among the nucleons, as they can be understood
within or follow from N. Bohr's compound nucleus.  Below we
shall try to elaborate on such a point of view on the basis of
our present day understanding of nuclear dynamics.  First we
want to explain why and in which way large scale motion may be
described by a linear response approach.

The basic element is the observation that motion in the
collective phase space may be expressed in terms of propagators
by writing for the time evolution of the density distribution
\bel{propagdens} f(Q,P,t) = \int {\rm d}Q_0 {\rm d}P_0 \; \; 
    K(Q,P,t;Q_0,P_0,t_0) \; \; f(Q_0,P_0,t_0) 
\ee
with
$ \lim_{t \to t_0}{K(Q,P,t;Q_0,P_0,t_0)}
    = \delta (Q-\Qtzer) \delta (P-\Ptzer) $
\begin{figure}[htb]
\centerline{\rotate[r]{\epsfysize=12.5cm \epsfxsize=8.cm 
\epsffile[104 58 549 669]{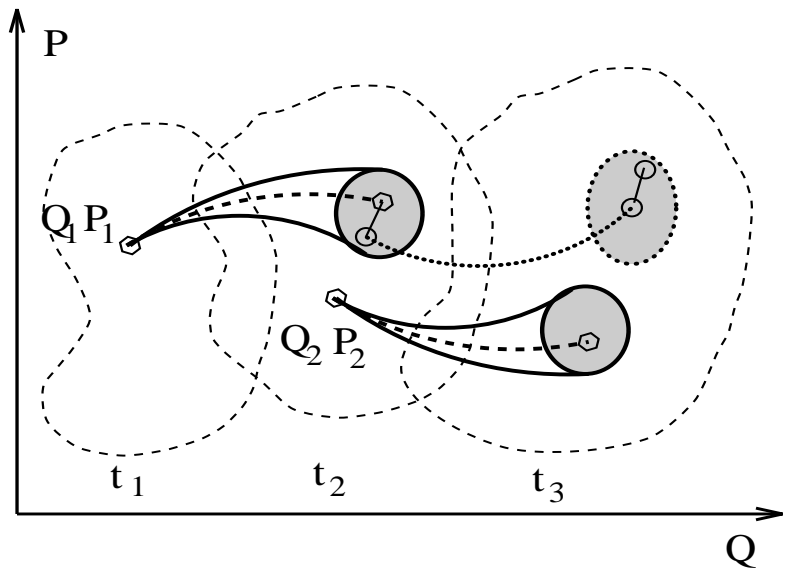}}}
\caption{\label{Fig.1}
Time development of global motion within the propagator
method. For the time step from $t_2$ to $t_3$ the related
Langevin method is indicated.}
\end{figure}

Here $K(Q,P,t;Q_0,P_0,t_0)$ may be interpreted as the conditional
probability for the system to move from $Q_0,P_0$ at $t_0$ to 
$Q,P$ at time $t$. On both sides of this relation the distribution $f$,
the "joint probability", may be replaced by conditional probabilities
defining the transition say from a $t_0$ to the final time $t$
through an intermediate step at $t_1$. The resulting relation is
nothing else but the Chapman-Kolmogorov equation.  (For a
discussion of such general properties we may refer to the book by van
Kampen \cite{vankampen}). 
\begin{figure}[htb]
\centerline{\rotate[r]{\epsfysize=12.5cm \epsfxsize=8.cm 
\epsffile[104 58 549 669]{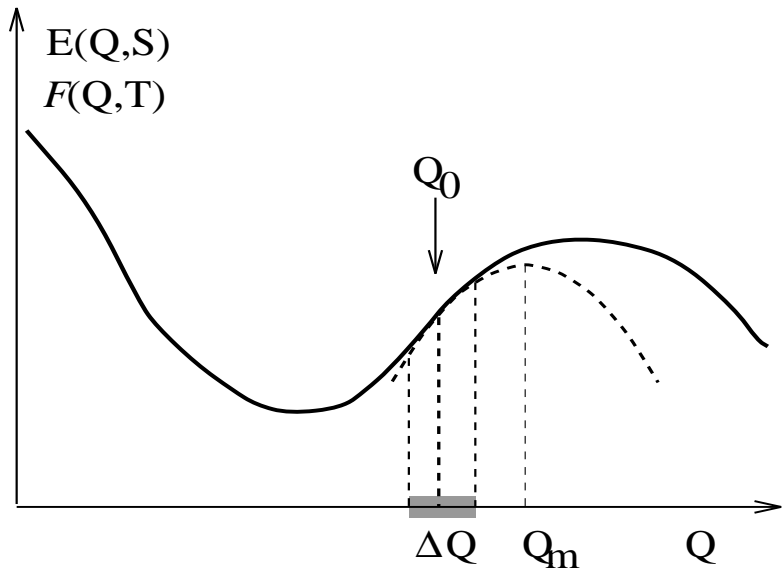}}}
\caption{\label{Fig.2}
The LHA to a typical fission potential, which may either
be given by the internal or the free energy.}
\end{figure}

The procedure just described may be repeated introducing small
time steps $\delta t$. They only must be chosen longer than the
typical microscopic time $\tau$ below which an averaging over
the microscopic dynamics of the nucleons would become
meaningless. A linear response approach becomes possible
whenever the $\delta t$ is {\it small on the collective time
scale}. Consequently, the $K(Q,P,t;Q_0,P_0,t_0)$ may be
constructed within a {\it locally harmonic approximation (LHA)},
to collective motion.  If this propagator spreads over a limited
regime of phase space only, within the time lap $\delta t$, a
Gaussian form will do. A pictorial view of this method is given
in Figs.1 and 2.

\subsection{The equation of motion for the propagator}

To approximate the propagator by a Gaussian means to write:
\bel{gaussprop} K_G(Q,P,t;Q_0,P_0,t_0) = (2\pi)^{-1}{\it \Delta}^{-1/2}\; 
    \exp\Bigl\{-{1\over 2}\sum_{i,k}({\cal X}_i-{\cal X}_i^c(t))
    ({\cal X}_k-{\cal X}_k^c(t)) {\cal A}_{ik}(t)\Bigr\}      
\ee
with the ${\it \Delta}(t)$ being the determinant
${\it \Delta}(t) = \det \bigl({\cal A}^{-1}_{ik}(t)\bigr)$.
The vector ${\cal X}$ defines a point in collective phase space,
and the $K_G$ represents a distribution, centered at the
trajectory ${{\cal X}^c(t)}$, which starts at $P_0, Q_0$,
\bel{phasespvec} {\cal X} =\pmatrix{ P \cr
                                  Q \cr} \qquad 
       {{\cal X}^c(t)} = \pmatrix{ P^c(t) \cr
                                  Q^c(t) \cr} 
                           = \pmatrix{ \bra P \ket_t \cr
                                  \bra Q\ket_t \cr} 
  \qquad \pmatrix{ P_0 \cr  Q_0 \cr}  =\pmatrix{ \Ptzer \cr
                          \Qtzer \cr}
\ee
and which fulfills the following set of equations:
\bel{diffeqqp}\pct =M{dq^c\over dt} \qquad\qquad
    {dP \over dt}=-C q^c(t)  -\gamma {dq^c\over dt} 
\ee
Here we have introduced the variable $q = Q-Q_m$.
The $Q_m$ defines the position of the local oscillator, more precisely
the position of the extremum of the local, effective collective
potential. It is a linear function of $\Qtzer$, and has to be defined by the
condition that the constant term in the force disappears see
Fig.2 and below.

The matrix ${\cal A}_{ik}(t)$ is given by the widths of the
distribution 
\bel{matfsm} \bigl({\cal A}(t)\bigr)^{-1} =
    \pmatrix{ \sigmapp & \sigmaqp \cr
              \sigmaqp & \sigmaqq \cr} 
\ee
with the $\sigmaqq $ etc. being defined as 
$\sigmaqq =\bra(Q-\qct)^2\ket_t \equiv \bra
    Q^2\ket_t-(\qct)^2 $. 
These second moments fulfill the following set of
equations:
\bel{marksigqq} {d \over dt} \sigmaqq  - {2\over M}
   \sigmaqp = 0
\ee
\bel{marksigqp} {d \over dt} \sigmaqp  - {1\over M} \sigmapp + 
 C \sigmaqq + {\gamma \over M} \sigmaqp = D_{qp} 
\ee
\bel{marksigpp} {d \over dt} \sigmapp  + 2C \sigmaqp 
+ 2{\gamma \over M} \sigmapp =2D_{pp} 
\ee
The inhomogeneities represent the diffusion coefficients.
The initial condition
$ \sigmaqqini =\sigmaqpini = \sigmappini = 0$
warrants the $K_G$ to start from the sharp distribution as
required by (\ref{propagdens}). 

This $K_G$ satisfies the following transport equation
\bel{traneq}{\partial  \over \partial t} f(q,P,t) = \Biggl[
    -{\partial\over\partial q}{P\over M} + {\partial\over\partial P}Cq
    +{\partial\over\partial P}{P\over M}\gamma +
     D_{qp}{\partial^2\over\partial q\partial P} + D_{pp}
    {\partial^2\over\partial P\partial P} \Biggr] f(q,P,t) 
\ee
This is easily verified by mere differentiation. The form of
this equation resembles the one of Kramers with the exception
that in his case the diffusion coefficients were given by
$ D_{pp} = \gamma T$ and $ D_{qp} = 0$.
Moreover, as he understood his equation to describe large scale
motion, the conservative force was given by the derivative of
the full potential, $-\partial V(q) /\partial q$, rather than by $-Cq$.

\section{The derivation of the EOM within the LHA}

In this chapter we are going to describe briefly how we may obtain
the transport coefficients as they appear in (\ref{traneq}), or in
the equations for the first (\ref{diffeqqp}) and second moments
(\ref{marksigqq}-\ref{marksigpp}). To this end let us assume to be given
a Q-dependent Hamiltonian of the type 
$\hat H(\hat x_i,\hat p_i,Q)=\hat H_{sp}(Q) + \Vres $
The term $\hat H_{sp}(Q)$ represents independent particle motion 
in a deformed shell model, with Strutinsky renormalization meant to be
included. The $\Vres$ stands for a residual interactions, considered
finally by way of a complex self-energies of the single particles:
$ \self(\om\pm i \epsilon,T) = \selfre(\om,T) \mp {i\over 2} \Gamma(\om,T) $.
Microscopic expressions are then calculated after replacing the
single particle strength 
$\varrho_k(\om)= 2\pi\;\delta(\hbar \om - e_k) $ by
\bel{dspspecdensgam} \varrho_k = {\Gamma(\om) \over
    \left(\hbar \om -e_k -\selfre(\om)\right)^2 + 
    \left({\Gamma(\om)\over 2} \right)^2}
\qquad \Gamma=
    {1\over \Gamma_0}\;{(\hbo - \mu)^2 + \pi^2 T^2 \over 
    1 + \left[(\hbo - \mu)^2 + \pi^2 T^2 \right] /c^2}
\ee
with the $\mu$ being the chemical potential. The $\Gamma$ is understood
to represent "collisional damping". In numerical computations,
the following values have mostly been used for the parameters entering here: 
$\Gamma_0 = 33 \mev $ and $ c= 20 \mev$.

\subsection{Average motion}

\subsubsection{Secular equation from energy conservation}

An equation of average motion may be obtained from energy
conservation \cite{holet}, regarding the nucleus as an {\it
isolated system}. One may thus write
$ 0 = {d \over dt} E_{\rm tot} = {\dot Q}
   \left<{\partial \ham{Q} \over \partial Q}\right>_{t} \equiv 
    {\dot Q} \left< \ffield{Q} \right>_{t}
$,
to get the equation of motion for $Q(t)$ after expressing
the average $ \left< \ffield{Q} \right>_{t}$ as a functional of
$Q(t)$. Following the scheme of the LHA one may
expand the $\hat H(Q)$ around any given $Q_0$ to have:
\bel{hamilapp}\hat H(Q(t))=\hat H(Q_0)  + 
     (Q(t)-Q_0)\h{F} +
     {1\over 2}(Q(t)-Q_0)^2 \dtwohqzero   
\ee
The effects of the coupling term
$(Q(t)-Q_0)\h{F}$ may then be treated by linear response theory,
exploiting as a powerful tool the causal response function $\rft$ 
\bel{defrest} \rft (t-s)= \Theta(t-s) {i\over \hbar} 
             {\rm tr}\,\left(\hat{\rho}_{\rm qs}(Q_{0},T_{0})
             [\hat F^{I}(t),\hat F^{I}(s)] \right) 
     \equiv 2 i \Theta(t-s) \drft(t-s)
\ee
Here, the time evolution in  $\hat F^{I}(t)$ as well as
the density operator $\equiop$ are determined by $H(Q_{0})$. The
$\equiop$ is meant to represent thermal equilibrium at $Q_0$ with
excitation being parameterized by temperature $T_0$ or by entropy $S_0$.

The equation of motion may finally be brought to the form
$k^{-1}q(t) +
    \int_{-\infty}^{\infty}\widetilde{\chi}(t-s) q(s) {\rm d}s =0 
$
whose Fourier transform leads to the well known form
of the  {\it secular
equation} for the possible local frequencies of the harmonic
motion:
$ k^{-1} + \chi(\om) = 0$. Its solution will be discussed below.
The $q(t)$ is given by
\bel{q}q(t) = Q(t)-Q_m \qquad {\rm with} \qquad
    \left.{\partial E(Q,S_{0})\over \partial Q}\right\vert_{Q_{0}}
   + \left.{\partial^{2} E(Q,S_{0})\over \partial Q^2}\right\vert_{Q_{0}}
             (Q_{m}-Q_{0}) = 0 
\ee
and $Q_{m}$ represents the position of the center of the local
oscillator and the $k$ a coupling constant, which for the
case of zero temperature is known from the Copenhagen version of
describing harmonic vibrations \cite{bm}. At finite thermal
excitations things become more complicated, where besides the
shape variable itself also the parameter specifying the state of
the system may change in time. Choosing entropy, however, one
may argue this change to happen only in second order in the
velocity. For harmonic motion this is beyond the order to be
considered. Keeping $S$ constant one then obtains \cite{kidhofiva}
\bel{krik}-k^{-1} = \dtwohqzero + (\chi(0)-\chi^{\rm ad})
    =
   \left.{\partial^{2}E(Q,S_{0})\over \partial  Q^{2}}\right\vert_{Q_{0}}
   +\chi (0)
\ee
The
$\chi ^{\rm ad} = -\left({\partial <\hat F>\over\partial
Q_0}\right)_{S} $
is the {\it adiabatic susceptibility} and $E(Q,S_0)$ is the internal
energy. As the latter is somewhat 
difficult to evaluate one may introduce the free energy to rewrite
(\ref{krik}) as 
\bel{kreal} -k^{-1} = 
    \left.{\partial^{2} \totfren(Q,T_0)\over \partial Q^{2}}\right\vert_{Q_{0}} + 
    \chi(0) -\left({1\over {\partial^{2}\totfren\over \partial T^{2}}} 
    \left({\partial^{2}\totfren\over \partial T\partial Q}\right)^{2}
    \right)_{Q_{0},T_{0}}
\ee
Notice, please, that for constant entropy $S$ one may, for the
quasi-static case, relate the temperature change to the one of
the collective coordinate itself by applying general rules from
thermodynamics: $dT = \left({\partial T / \partial Q}\right)_{S}
dQ $.

Another point which we want to elaborate further below concerns the
difference  between the adiabatic susceptibility and the static
response, which can be shown to be either positive or zero. It vanishes
at zero thermal excitations but at finite temperature it does so only
for {\it ergodic systems}, a problem to which we shall return below:
\bel{conderg} \chi^{\rm ad} - \chi(0) = 0
\ee

\subsubsection{ The collective response}

Let us introduce a coupling  $q_{ext}F$ to a
time dependent "external field" $\fext(t)$
by adding  $\fext(t) \hat F$ to our Hamiltonian:
$ \hat H^\prime=\hat H(\xvi,\pvi,Q) + \fext(t)\hat F $
but where $\hat H(Q)$ is taken in the approximate form (\ref{hamilapp}).
The coupling is chosen to have the same form as the one between the
two "subsystems" of collective and nucleonic degrees of freedom, with
only the $(Q(t)-Q_0)$ replaced by the "external field" $\fext$.
The "collective" response function may now be defined as:
\bel{defcollres} \delta \fmb_\om   = - \rfc{\om} \fext(\om) 
\ee
where for convenience we have switched to frequency representation. The
difference between the $ \delta \fmb_\om$ and the $\fmb_\om$ should at best
have a term proportional to $\delta(\om)$, representing some static
force in the time dependent picture. The phrase "collective" response
function is chosen because i) this function now includes the collective
excitations as well and ii) we will later on consider the latter as the
only interesting ones.

The construction of the $\respc$ can be done following the
derivation of the polarizability function for electric media as
introduced by Clausius and Mosotti, and as described in \cite{bm}  for
the nuclear case at zero excitation. We need three equations: Besides
the basic definition (\ref{defcollres}), we have to have an equation which
relates $\fav{t}$ to $(Q(t)-Q_0)$ and we need to express $\fav{t}$ by
the total perturbation consisting of both {\it external and induced fields}. 
We want to perform this construction assuming ergodicity for our nucleonic
system, i.e. requiring the condition (\ref{conderg}) to be fulfilled.
Exploiting (\ref{krik}) the relation between the average $\fav{t}$ 
and the deviation of $Q$ from $Q_0$ becomes
\bel{fexvald} k \fav{t}= Q-Q_{0} 
\ee
Finally, the collective response function defined
by (\ref{defcollres}) takes on the form
\bel{rfcoll} 
    \rfc{\om} = {\chi(\om)\over 1+k\chi(\om)}
\ee

The relation (\ref{fexvald}) reminds one of a {\it self-consistency} condition, like 
one would find treating an effective two-body interaction $(k/2) {\hat
F}{\hat F}$ in mean field approximation. However, such a notion is
justified on more general grounds. It is this relation (\ref{fexvald})
which expresses most clearly that the correct interpretation of the
$Q(t)$ must be to consider this quantity an {\it internal}
variable---measuring some properties of the nucleus and clearly to be
distinguished from truly external fields like the $\fext$.

\subsubsection{Transport Coefficients and
the distribution of collective strength}

The procedure of {\it defining} the transport coefficient of
inertia, friction and local stiffness is as follows:  
For {\it selected peaks} of the strength distribution one fits
the dissipative part of the oscillator response $\chiosc
 =  \left( M\om^2 + i \gamma \om -C\right)^{-1} $ to
the $\rfc{\om}$ calculated microscopically.  If one is interested
in the transport coefficients for the $q-$motion one first
transforms to the response function $\resqq$ in the $q$-mode,
defined through $\resqq = {\delta q(\om) / \delta (-\qext(\om)}
\mid_{\qext=0} $. This can easily be done with the help of the
linear relation: $ k \delta \fav{t}= \delta Q(t) \equiv  q(t) $
which implies $ \resqq = k^2 \rfc{\om}$. Summarizing, this fit
then implies the substitution (with $\om_1^{\pm} = \pm {\cal E} -
i \Gamma/ 2$)
\bel{restonemo}\resqq\; \Longrightarrow \; \chiosc=
 {-1 \over M(\om_1^{+}-\om_1^{-})} \left( {1 \over
    \om - \om_1^{+}} -{1 \over \om -\om _1^{-}}\right) 
\ee
in a certain range frequencies which encompasses the peak in the
strength distribution whose associated mode $\om_1$ one wants to treat
explicitly. Notice, please, that in this way we actually solve the 
secular equation for arbitrary damping. In the case of fission it is
believed that it is mainly the low frequency behavior which matters,
for which reason it is the lowest peak which is
treated in this way. For an application to giant resonances see
\cite{hoyaje} for the case of the quadrupole and the talk by S. Yamaji at
this meeting for monopole vibrations.

Within the formulation presented above 
the strength distribution may change dramatically with temperature
\cite{hoyaje}, \cite{hoivyanp}, \cite{yaivho}. This is mainly due to the temperature
dependence of the coupling constant, as determined by the stiffness
$C(0) = \partial^{2}E(Q,S_{0})/\partial  Q^{2}$ of the static energy
(see (\ref{krik})). For the quadrupole as well as for the local motion
along the fission path, the $C(0)$ may well decrease by more than an
order of magnitude when $T$ is increased from $0$ to about $2\cdots 3
\mev$. As a consequence, at the higher temperatures practically {\it
all strength} concentrates in {\it one} dominant low frequency mode,
with the associated transport coefficients reflecting strongly {\it
over-damped motion}. This feature has been demonstrated in \cite{hoyaje}
(see Fig..1 of this reference or Fig.5.1.1 of \cite{hofrep}). 
As may be expected from such a strength distribution, the inertia of
the low frequency mode decreases from the typical cranking model value
at low excitations to the one given by the energy weighted sum rule at
large $T$ \cite{hoyaje}.

Conveniently the transport coefficients are expressed by the following
ratios 
\bel{defgameta}{\Gamma } = {\gamma \over M}  
    \;\;\;\quad \;\;\;\;\;\;\quad \;\;\;
    \varpi ^2= {{\mid C \mid }\over M}
    \;\;\;\quad \;\;\;\;\;\;\quad \;\;\; 
     \eta  = {\gamma \over 2\sqrt{M\mid C \mid}} 
\ee
with which the frequencies the two frequencies may be expressed as 
$ \om ^{\pm } = \pm  {\cal E} -i {\Gamma \over 2} 
    =\varpi \left( \pm \sqrt{{\rm sign}C - \eta^2} - i\eta \right) $.
Numerical values from microscopic computations can be found in
\cite{hoivyanp}, \cite{yaivho} and \cite{ivhopaya}. A typical
calculation of $\Gamma$ is shown in Fig.3 (from
\cite{ivhopaya}). If evaluated at the
potential minimum and at the barrier, typically the $\hbar\varpi$ is of
the order of $1\mev$, independently of $T$; conversely the $\eta$
increases from about $1$ at $T=1\mev$ to about $3$ at $5\mev$
\cite{ivhopaya}.  

\begin{figure}[h]
\centerline{{
\epsfysize=8cm
\leavevmode
\epsffile[111 286 533 544]{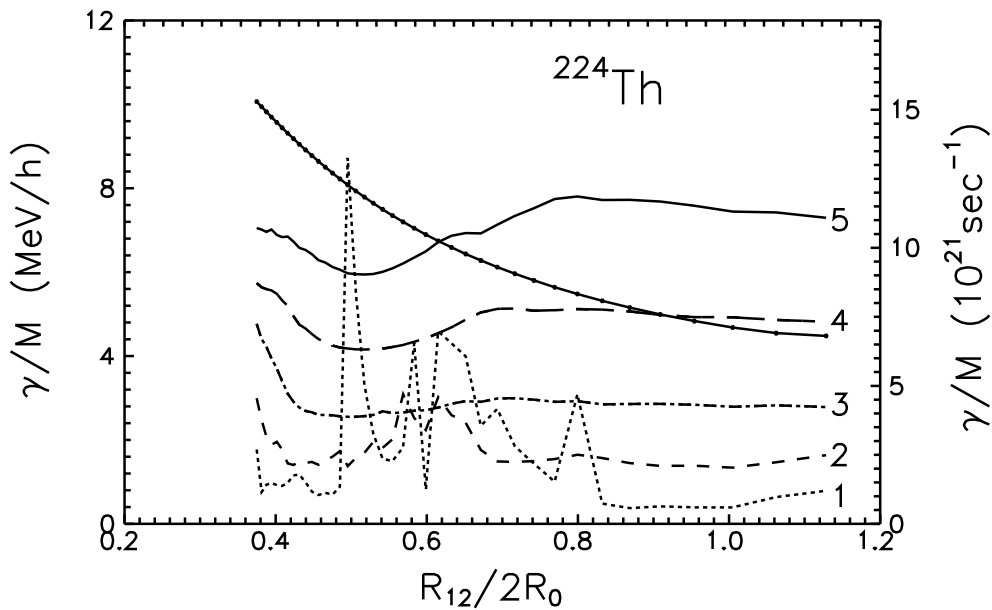}
}}
\caption{The inverse relaxation time $\Gamma =\gamma 
(\omega_1)/M(\omega_1)$
as function of deformation and temperature (indicated in the Figure).}
\label{Fig.3}
\end{figure}

The behavior found in our model for the strength distribution may be
understood in the sense of reaching a {\it macroscopic limit} with
increasing $T$. Within such a limit there is no room any more neither
for quantal coherence in the sense of collective motion nor for the
typical single particle excitations. The nucleus just behaves like a
drop of a strongly viscous nuclear fluid (see also \cite{him}, \cite{maghof}).
This behavior is in clear contrast to the one found for common RPA at
finite $T$, for instance like in the so called "thermo field theory"
(see talks at this meeting and c.f. \cite{khqstsu}).

\subsubsection{The equation of average motion and the
transfer of heat}

After the substitution (\ref{restonemo})
the equation for the $q$-mode attains the form:
\bel{diffeqq}  M(\om{_1})\, {\ddot q}(t) + 
    \gamma(\om{_1})\,{\dot q}(t) +C(\om{_1})\, q(t) = -\qext (t)  
\ee
It is of
{\it differential} nature although we have {\it not} really made use
of any {\it Markovian approximation}. All we did is to {\it restrict}
ourselves to just {\it one mode} assuming that the {\it associated peak
can be approximated by a Lorentzian}. For these reasons we are actually
dealing with {\it Non-Markovian transport coefficients} representing
dynamics in a {\it whole range of frequencies}. 
They are determined  {\it uniquely} for {\it under-} as well as for
{\it over-damped} motion, and for {\it positive or negative stiffness}.

Let us turn to the {\it energy balance} once more. One finds the equation
\bel{enerbala}{d\over dt}E_{\rm tot}={d\over dt}\left({M{\dot q}^2\over 2}+
{Cq^2\over 2}\right) + \gamma \dot q^2
={d\over dt}E_{\rm coll}+T{dS\over dt} 
\ee
no matter whether the external field is "switched on or off"
\cite{kidhofiva}. For (damped) self-sustained motion the equation $
{d E_{\rm tot}/dt}  =0$ may thus be rewritten as
\bel{decolldiff}-{d\over dt} E_{\rm coll}  \equiv
    -{d\over dt}\left({M(\om_{1})\over 2}{\dot q}^{2}
    +{C(\om_{1})\over 2}q^{2}\right) = \gamma(\om_{1}){\dot q}^{2}
    \equiv  T {d\over dt}S 
\ee
which correctly expresses the exchange between collective motion into
heat.

\subsection{The dynamics of fluctuations of the collective
degrees of freedom}
So far the collective variable $Q$ was just a c-number classifying the
deformation of the {\it mean field}. To be able studying collective
fluctuations we first must raise $Q$ to become a {\it genuine}
dynamical variable. As we are interested in quantum effects this
generalization has to be performed in {\it operator sense}.
Simultaneously, we need to introduce the {\it associated momentum} 
$\hat \Pi $, and we must find a {\it decent Hamiltonian for the total
system}. Such a Hamiltonian may then be used to derive the transport
equation for the collective density operator $\hat d$, the Wigner
function of which may be associated with the distribution in collective
phase space mentioned before. Evidently, such a procedure ought to
account for the basic and important features we have been able to
consider for average motion, the most stringent one perhaps being
self-consistency. If at all possible, this ambitious program can be
accomplished only in a restrictive sense; here it will be the LHA
approximation, once more, which is exploited.

\subsubsection{A suitable Hamiltonian for the total system}

We would like to have a Hamiltonian which splits into
parts of nucleonic and collective motion plus a coupling between both
subsystems: $ \hat {\cal H}=\hat H_{\rm nucl}+\hat H_{\rm coup}+\hat H_{\rm
  coll} $. This goal may be achieved by applying the Bohm-Pines
method, which originally has been invented to treat collective modes in
the electron gas. One obtains: 
\bel{bpham}
\hat {\cal H}
  =\hat H(\hat x_i,\hat p_i,Q_0) + 
  k \hat \Pi \hat {\dot F}-{\beta \over k} (\hat Q - Q_0)\hat F +
   {\hat \Pi^2 \over 2 m_0 } + 
    {1 \over 2 k^2} (2 \beta + k) (\hat Q - Q_0)^2
\ee
The unperturbed collective inertia is given by the sum rule value, namely
\bel{coinun} {1 \over m_0} = i k^2 \; \langle \, [\hat {\dot F},\hat F ] \,
    \rangle _{Q_0} \qquad \qquad
 \hat {\dot F} = i \; [\hat H(Q_0) , \hat F] 
\ee
The $\hat \Pi$ is the canonical momentum satisfying the 
commutation rule $[\hat Q,\hat \Pi] = i \hbar$. 

The intrinsic part obtains a most convenient form: It is the same
$\hat H(\hat x_i,\hat p_i,Q_0)$ which appeared already in 
(\ref{hamilapp}) as the unperturbed part. The important point is that there
is {\it no reduction} of the number of nucleonic degrees of freedom.
This implies that we will be able to work with the same nucleonic
response functions as before. 
Of course, in order not to have too many degrees of freedom
there will have to be a subsidiary condition, which turns out to be
\bel{subcond} k\hat F - (\hat Q - Q_0) = 0 
\ee
For average motion it reduces to the form (\ref{fexvald})
found before. Notice the {\it two coupling terms}: The first one
is of similar structure as that in (\ref{hamilapp}). In addition there is
term involving $\hat P$, which is multiplied by the same factor
$\hat {\dot F}$ which appears also in the unperturbed inertia.  It is
this intimate relation between the coupling terms and the unperturbed
collective Hamiltonian which in the end assures self-consistency, in
the sense of the collective coordinate being related to the intrinsic
field $\hat F$.
The alert reader will have noticed that there is the additional
parameter $\beta$. In the end the latter is chosen as
$(\beta + k)/ k^2 = C = -M \omega_1^{+} \omega_1^{-}$,
Very naturally this goes along with the re-introduction of 
the kinetic momentum, such that for the averaged dynamics one regains the
form (\ref{diffeqqp}). 

For undamped motion the situation is like in common RPA: The
Hamiltonian can be transformed in such a way that the coupling
between collective and nucleonic degrees of freedom effectively
shows up in renormalized transport coefficients for inertia and
stiffness, after the intrinsic degrees of freedom have been
averaged out. The corresponding frequency satisfies the secular
equation of RPA which corresponds to the separable interaction
$(k/2) {\hat F}{\hat F}$.  This equation has the form given
above but where in the $\chi$ only the reactive part plays a
role. For damped motion the situation is more complicated. To
get information about the dynamics of collective fluctuations
one needs to invoke projection techniques, like the one of the
Nakajima-Zwanzig method.

\subsubsection{A non-perturbative Nakajima-Zwanzig
approach}

One starts with the von Neumann equation for the Hamiltonian (\ref{bpham})
and defines the reduced density $\hat d$ by averaging out the
intrinsic degrees of freedom. Its equation has a term which
is non-local in time. The integral kernel $\cal K$  contains an
operator for the time evolution of the total system, which commonly is
treated in second order perturbation theory, the argument being that
$\cal K$ does already contain the coupling twice. However, in order to
meet the standard set up at the beginning of this subsection one needs
to do better. The new scheme to be applied may be read of from
the structure of the collective response function (\ref{rfcoll}): On
the one hand one may exploit low order perturbation theory with
respect to the effects of the coupling on the intrinsic degrees
of freedom, for which reason the nucleonic response function
$\chi$ appears. Conversely, with respect to collective motion an
infinite order is considered, as seen by the fact that the
coupling constant appears in the denominator. To translate this
scheme to the NZ approach essentially one replaces the
unperturbed time evolution operator by an {\it effective} one:
\bel{modpernz}\exp[-i(\hat L_{\rm coll} + \hat L_{\rm nucl})s]
\quad \Longrightarrow \quad \exp[-i(\hat {\cal L}^{eff}_{\rm coll} +
\hat L_{\rm nucl}) s] 
\ee
Here, time evolution is formulated in terms of Liouvillians. Fortunately,
the effective one,  $\hat {\cal L}^{eff}_{\rm coll}$, need
not be constructed explicitly if we are interested only in the Gaussian
form of the local propagators introduced in sect.1. It suffices to exploit
the quantal fluctuation dissipation theorem, which can be proven to be
satisfied within the {\it modified Nakajima-Zwanzig approach}.

\subsubsection{Diffusion coefficients and the equilibrium of the
damped oscillator}

Let us turn our attention to equilibrium fluctuations first.  Imagine
the {\it total} system to be close to equilibrium at some given
temperature $T$.  The collective fluctuations may then be calculated by
applying the {\it quantal fluctuation dissipation theorem}, plus a few
general relations of response functions. For the three possible
combinations of $q$ and $P$ one gets:
\bel{dkinqqpp} D_{pp} = {\gamma \over M} \oscsigmapp
   \qquad D_{qq} = {2\over M}\eqsigmaqp = 0 
   \qquad D_{qp} = C \oscsigmaqq - {1 \over M}\oscsigmapp  
\ee
It is quite easy to convince one-selves that these relations are in
accord with (\ref{marksigqq}-\ref{marksigpp}) for stationary situations. The
equilibrium fluctuations are determined by
\bel{oscaverqqpp} \oscsigmaqq =\int \; {{\rm d}\omega \over 2\pi }\, 
  \hbar\coth \left( {\hbar\omega \over 2T}\right) \, \respposc 
  \qquad 
  \oscsigmapp = M^2 \,\int \; {{\rm d}\omega \over 2\pi }\, 
  \hbar\coth \left( {\hbar\omega \over 2T}\right) \, \omega ^2\respposc
\ee
For the kinetic momentum we have used the fact
that $P=M\dot q$. The cross fluctuation vanishes
because $q$ and $P$ behave differently under time reversal.
We are now going to demonstrate first that for two limiting cases these
expressions lead to results anticipated on general grounds.

{\it (i)  Limit of high temperature:}
Let us begin addressing a situation for which we expect the
Einstein relation proper to apply. Suppose the temperature to be
sufficiently high such that in the integrals appearing in (\ref{oscaverqqpp})
we may effectively replace $\hbar\coth({\hbar\omega / 2T})$ by 
$2T / \om $.
Then for the average kinetic and potential
energy we just obtain the values known from the equipartition theorem, namely
\bel{equipartht} {\eqsigmapp \over 2M}= {\langle \; \hat P^2 \; 
  \rangle _{eq} \over 2M} = {T\over 2} =
   {C\over 2}\langle \; \hat q^2 \; \rangle _{eq}
   = {C\over 2}\eqsigmaqq  
\ee
For the diffusion coefficients given in
(\ref{dkinqqpp}) this indeed implies the forms given below eq.(\ref{traneq}).

{\it (ii) Limit of zero damping:}
Simple results can also be obtained for the limit of zero friction.
Take the oscillator response and let the imaginary part
$\Gamma $ of $\om_1^\pm$ go to zero. Then the dissipative part of the
response function can be written as
$\lim_{\gamma \to 0} {\chi^{\dpr }_{\rm osc} (\omega )} = 
  (\pi/ 2M\varpi) \left( \delta (\omega -   \varpi)- \delta 
  (\omega + \varpi)  \right) $.
For such a mode, the average values of the kinetic and potential
energies take on values given by the quantal version of the
equipartition theorem, namely:
\bel{equipartzd} {\eqsigmapp \over 2M}= {\langle \; \hat P^2 \; 
  \rangle _{eq} \over 2M} =  {\hbar \varpi \over 2}
  \coth ({\hbar \varpi \over 2T}) =
  {C\over 2}\langle \; \hat q^2 \; 
  \rangle _{eq}  = {C\over 2}\eqsigmaqq  
\ee
Also, the diffusion coefficients then look like the ones obtained before
with only $T$ replaced by an effective temperature $T^*(\varpi)$,
namely 
\bel{efftemp}  D_{qp} = 0
   \qquad  D_{pp} = \gamma T^*(\varpi) 
   \qquad {\rm with }\qquad 
   T^*(\varpi ) = {\hbar \varpi \over 2}
  \coth ({\hbar \varpi \over 2T}) 
\ee
This version of the generalized Einstein relation has been used in
ref.\cite{hogngo} to describe the "fast" mode of charge equilibrization in
heavy ion collisions.

{\it (iii) General case:}
The integrals in (\ref{oscaverqqpp}) 
may be evaluated in terms of infinite series by applying the residue
theorem, once the hyperbolic cotangent has been developed into the
uniformly convergent pole expansion involving the Matsubara
frequencies. For evaluating $\oscsigmapp$ the integral must be
regularized, for instance by introducing a frequency dependent
friction coefficient which drops to zero beyond the so called Drude
frequency $\om_D$, for details please see \cite{khqstsu}. Typical
numerical results are shown in Fig.4, where the "equilibrium
fluctuation"  $\eqsigmaqq$ is shown together with both diffusion
coefficients, all quantities normalized in self-explaining fashion.
They are plotted as function of the local stiffness $C$, which becomes
negative when the local modes become unstable. It can be proven that
this happens exactly when the (unperturbed) stiffness $C(0)$ of the
static energy turns negative. The figure demonstrates that our way of
calculating the diffusion coefficients is continuously possible
independent of the sign of the stiffness. It is only for smaller
temperatures that one runs into difficulties  for unstable modes; we
will address this point later on. Incidentally, one also observes that
for $C<0$ the $\eqsigmaqq$ itself is negative. This does no harm as
for unstable modes this quantity itself has no physical meaning.
\begin{figure}[htb]
\centerline{\rotate[r]{\epsfysize=12.5cm \epsfxsize=8.cm 
\epsffile[104 58 549 669]{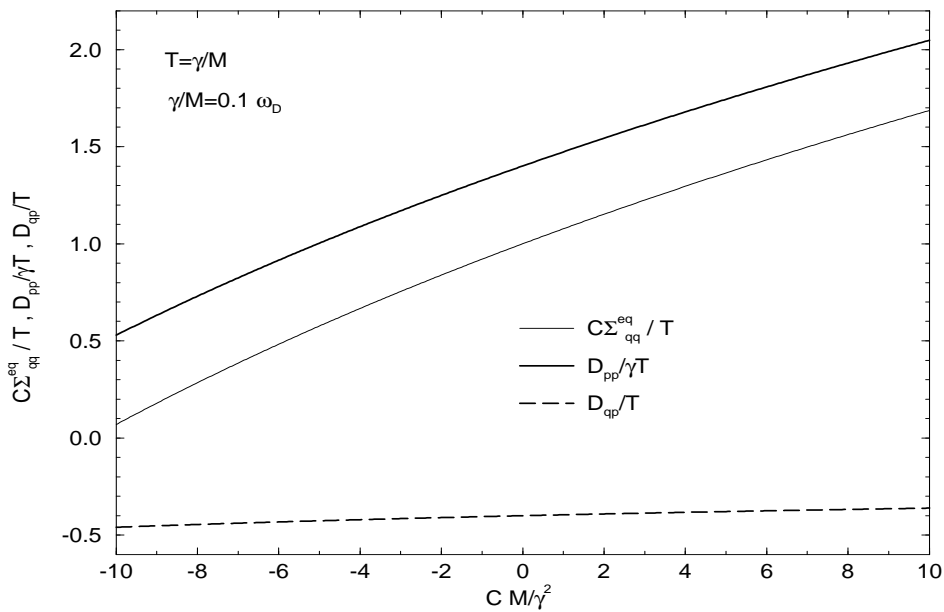}}}
\caption{\label{Fig.4}
Diffusion coefficients (diagonal: thick fully drawn line,
off diagonal: dashed line) and analytic continuation of the equilibrium
fluctuations in $q$ (thin full line) as functions of the stiffness $C$,
for fixed values of $M$ and $\gamma$. 
}
\end{figure}

\section{Open questions}

In this section we want to address a few of the most stringent but
unresolved problems; to some extent they have been mentioned already in
papers cited above. 

\subsection{Nuclear thermodynamics}
It has turned out unavoidable to benefit from the concepts of
thermodynamics when classifying excitations of nuclear systems,
knowing the latter not only to be small and isolated but to be unstable
as well at already such small values of overall excitations as 
$10 \mev$. Often, in the spirit of a shame-faced concealment, one uses
(or perhaps misuses) the excuse that temperature (or the total energy)
only parameterizes the level density. To me it is unclear what such a
phrase really means, considering the fact that even in thermodynamics
proper concepts like entropy and temperature only derive from certain
interpretations of the level density, even if applied to truly
macroscopic systems. Rather, one may apply the concepts of
thermodynamics in literal sense, if one only adds the question
about possible errors, in the quantitative sense.
Certainly, such a point of view pre-supposes that the system
equilibrizes at all. This question simply depends on time scale: There
must be sufficient time for most of the degrees of freedom available in
a certain process to envisage "all" its possible configurations before
the nucleus disintegrates in one way or other. Knowing that this
condition may not always be easy to fulfill, in particular at the somewhat
higher excitations, we want to start from such a hypothesis to concentrate
on the problem about the use of the concept of temperature or the
canonical distribution. Looking back at the various forms of the diffusion
coefficients this question certainly appears to be relevant also in the
more physical sense.

\subsubsection{Fluctuations of basic thermal quantities}

{\it (i) The canonical distribution:} Here one assumes the
temperature to have a definite value, fixed by the internal energy
$\men $. The latter, however, is not determined precisely but exhibits a
fluctuation which by way of the fundamental formula $\delmen  =  T^2
{\partial \men / \partial T} = T^2 C$ can be expressed through the
specific heat $C= T {\partial S / \partial T} ={\partial \men /
\partial T}$. The entropy $S$ is subject to the uncertainty $\Delta S= C$. 
To get the right order of magnitude of these quantities it is totally
legitimate to use the "Fermi gas" model for the level density: $
\staw(\men^*)  \propto \exp\left(2\sqrt{a \men^*}\right)$, with 
$S=2\sqrt{a \men^*}$ and $\men^* = \men - E_0 = a T^2$ being the
excitation energy. The specific heat becomes
$ C= 2 a T = S $ such that the relative fluctuations can be expressed
as 
\bel{flucenen}
 {1\over 2}\;{\delmen^*\over \men^*}= {\Delta S\over S}= {1\over \sqrt{C}}
 \qquad = {2\over \sqrt{AT}} = 
 \left({2{\rm MeV}\over A\men^*}\right)^{1/4} \qquad \left(a={A/ 8} \right)
\ee
As indicated on the very right, a level density parameter has been used
which is about twice as large as that for the genuine Fermi gas
(free particles in a box). In this sense our estimate may be
considered realistic for the nuclear case. From the typical values given
below in the Table.1 it is seen that the energy fluctuations take on
considerable values!

{\it (ii) The micro-canonical distribution}, of course, is exempt
from this deficiency. It may be defined as $\rho(e) = 1/ G$ for
$E-\Delta \leq e\leq E$, with $G$ measuring the number of levels
inside the the interval considered and with $\rho(e) = 0$
outside. As can be seen from Table.1, even for a very small
width $\Delta$ of the distribution there are so many states that
a statistical treatment is unavoidable. Temperature may again be
introduced through the relation $T^{-1}= \partial S(e) /
\partial e\mid_\men$, but now this quantity exhibits a finite
fluctuation whose relative value is given by
\bel{flucT} {\Delta T\over T}  = {1\over \sqrt{C}} =
 {1\over 2}\;{\delmen^*\over \men^*}  
\ee
At smaller excitations this ratio may become uncomfortably large, a
fact to which attention has been drawn already in \cite{feshbach}. On the
other hand, for an application to transport theory such a fluctuation
does not hurt too much as long as the transport coefficients will not
depend on temperature too sensitively. Such a situation may be
given at $T$ not smaller than $1\mev$
\bigskip
\leftline{Table.1}
a) Dependence on $A$ for $\men^*=20 \mev, \Delta = 0.01
\mev$ 
\medskip
\offinterlineskip \tabskip=0pt
 \halign{ \strut
          \vrule#&
          \quad 
          \hfil #   \quad &
          \vrule#&
          \quad 
          \hfil #   \quad &
          \vrule#&
          \quad 
          \hfil #   \quad &
          \vrule#&
          \quad 
          \hfil #   \quad &
          \vrule#&
          \quad 
          \hfil #   \quad &
          \vrule#          \cr
  \noalign{\hrule}
& A                   && $50$       && $100$       && $150$       && $200$      & \cr
  \noalign{\hrule}
& T                   && $1.8$      && $1.3$       && $1.0$       && $0.9$      & \cr
  \noalign{\hrule}
&$\delmen^* /\men^*$      && $0.42$     && $0.36$      && $0.32$      && $0.30$      & \cr 
  \noalign{\hrule}
&$\Delta /\men^*$ && $5\cdot 10^{-4}$     && $5\cdot 10^{-4}$      && $5\cdot 10^{-4}$      && $5\cdot 10^{-4}$      & \cr   
  \noalign{\hrule}
&$G(\men^*)$      && $3.7\cdot 10^5$     && $3.9\cdot 10^9$      && $4.8\cdot 10^{12}$      && $1.9\cdot 10^{15}$      & \cr   
  \noalign{\hrule}
}
\vskip 0.5cm
b) Dependence on $\men^*$ for $A=100, \Delta = 0.001
\mev$ 
\medskip
{\offinterlineskip \tabskip=0pt
 \halign{ \strut
          \vrule#&
          \quad 
          \hfil #   \quad &
          \vrule#&
          \quad 
          \hfil #   \quad &
          \vrule#&
          \quad 
          \hfil #   \quad &
          \vrule#&
          \quad 
          \hfil #   \quad &
          \vrule#&
          \quad 
          \hfil #   \quad &
          \vrule#          \cr
  \noalign{\hrule}
& $\men^*$              && $25$       && $75$       && $125$       && $175$      & \cr
  \noalign{\hrule}
& T                   && $1.4$      && $2.5$       && $3.2$       && $3.7$      & \cr
  \noalign{\hrule}
&$\delmen^* /\men^*$      && $0.34$     && $0.26$      && $0.22$      && $0.21$      & \cr 
  \noalign{\hrule}
&$\Delta /\men^*$ && $4\cdot 10^{-5}$     && $1.3\cdot 10^{-5}$      && $8\cdot 10^{-6}$      && $5.7\cdot 10^{-6}$      & \cr   
  \noalign{\hrule}
&$G(\men^*)$      && $1.3\cdot 10^{10}$     && $7.5\cdot 10^{20}$      && $2.5\cdot 10^{28}$      && $3.4\cdot 10^{34}$      & \cr   
  \noalign{\hrule}
}
\normalbaselines
\baselineskip 14pt plus 1pt

\bigskip

\subsubsection{The variation of temperature with deformation}

Applying transport equations to nuclear collective motion it is evident
that temperature cannot be treated as being constant (see e.g.
\cite{hosia}, \cite{holet}). One reason why $T$ may change is because of the
production of heat due to dissipation (as given by eq.(\ref{decolldiff})).
In practical applications this effect has commonly been taken into
account (for an early treatment see \cite{hofngo},\cite{ngohof}).
However, for an isolated system temperature may change even
without dissipation. Within our formulation this effect is
hidden in the way the coupling constant is calculated, and it is
this feature which we are going to address now.

In the hypothetical case of a {\it  constant} temperature the 
coupling constant would be given by a form similar to (\ref{krik}),
but with $S$ replaced by $T$, the internal energy $E$ replaced
by the free energy $\totfren$ and the adiabatic susceptibility by
the isothermal one $ \chi^{\rm T} =  -\left({\partial \fmb^{\rm qs}/ \partial Q}\right)_{T}
     \vert_{Q=Q_{0}}  $, namely:
\bel{ktempcon} -\left.k^{-1}\right\vert_{T={\rm const.}} 
    = \dtwohqzero + (\chi(0)-\chi^{\rm T})
    = \left.{\partial^2 \totfren(Q,T_0)\over \partial Q^2}\right\vert_{Q_0} 
    + \chi(0)     
\ee
For such a process the quasi-static entropy would change as
function of time. Therefore, if we believe in our 
argument given before, this version (\ref{ktempcon}) cannot be
expected to correctly represent the physical situation. However,
in many cases the difference is not very large. It can be
expressed by the difference between the isothermal and adiabatic
susceptibility which in turn may be calculated from derivatives
of the free energy:
\bel{dchtad} -k^{-1}+\left.k^{-1}\right\vert_{T={\rm const.}} =
    \chi^{\rm T}-\chi^{\rm ad} = -
    \left({1\over {\partial^{2}\totfren\over \partial T^{2}}} 
    \left({\partial^{2}\totfren\over \partial T\partial Q}\right)^{2}
    \right)_{Q_{0},T_{0}}
\ee
Indeed, usually the term on the right hand side is not very big.
Generally speaking, for $T\geq \simeq 1.5 \mev$ one finds
\cite{kiderlen}
$\chi^{\rm T}-\chi^{\rm ad}  \ll  \left.{\partial^2
\totfren(Q,T_0) / \partial Q^2}\right\vert_{Q_0T_0} $, perhaps except in
the neighborhood of turning points of the potential energy.
At small $T$, however, the assumption of ${T={\rm const.}}$ may
become questionable, safe for very special cases like
vibrations about the ground state minimum of nuclei with doubly
closed shells, whose deformation does not change with excitation.

\subsection{Ergodicity problem and the heat pole}
Above it had been mentioned that for ergodic systems the adiabatic
susceptibility becomes identical to the static response: $\chi^{\rm
ad}=\chi(0)$. Indeed, as shown in \cite{kubo}, \cite{brenigtwo} 
(see also \cite{hofrep})
it suffices to have (i) a non-degenerate spectrum of the intrinsic
eigenstates as well as (ii) a narrow distribution in the occupation of
these states. Evidently, both conditions will be difficult to fulfill
within the {\it pure single particle picture and the canonical
ensemble}. This statement may be made quantitative by invoking 
the correlation function $\psi^{\dpr }(\om)$ which relates to the 
dissipative part $\chi^{\dpr }(\om)$ by way of the fluctuation
dissipation theorem:
$\hbar \drf(\om) = \tanh\left({\hbo/ 2T}\right) \psi^{\dpr}(\om)$. 
As the $\psi^{\dpr }(\om)$ derives from
an expression similar to the one given in (\ref{defrest}), but with the
commutator replaced by an anti commutator, a $\delta$ function type
singularity appears at $\om=0$: 
\bel{corrsingreg} \psi^{\dpr }(\om)= \psi^{0}
    2\pi \delta(\om) \; + \;_R\psi^{\dpr }(\om)
 =   T\Bigl(\chi^{\rm T} - \chi(0) \Bigr)
    2\pi \delta(\om) \; + \;_R\psi^{\dpr }(\om)
\ee
with the $\;_R\psi^{\dpr }(\om)$ being regular at $\om=0$. 
As written here, the pre-factor $\psi^{0}$ of this $\delta$ function
can be expressed by the difference of the isothermal susceptibility and
the static response. In the sequel we like to call this contribution
the "heat pole", with the $\psi^{0}$ being its residue.

If evaluated within the pure independent particle picture one gets
\bel{heatzeroipm}  \psi^{0}_{\rm ipm} 
   = \sum_k \vert F_{kk}\vert^2  n(e_k)(1-n(e_k))  
  = T\,\sum_{k}\left\arrowvert{\partial n(e) \over 
     \partial e}\right\arrowvert_{e=e_{k}} 
     \left({\partial (e_k -\mu) \over \partial Q}\right)^2  
\ee
In \cite{hoivyanp} this $ \psi^{0}_{\rm ipm} $ has been calculated as
function of $T$. At low $T$ it increases rapidly to reach
a constant value above about $T=2\mev$. This value overshoots
considerably the one it would have for an ergodic system, namely
$\chi^{\rm T}- \chi^{\rm ad}$. Unfortunately, this situation does not
change when the collisional damping is taken into account. In this case
the heat pole attains a finite width
\bel{heatplor}  \corrheat(\om)= \psi^0 2\pi \delta(\om) \qquad
    \Longrightarrow \qquad \corrheat(\om) = \psi^0 {\hbar \Gamma_T
    \over (\hbar \om)^2 +
   \Gamma_T^2 /4}    
\ee
but the strength of the Lorentzian essentially remains unchanged, see
\cite{hoivyanp} and \cite{hofrep}. (It may be noted in
passing that the width $\Gamma_T$ increases almost linearly in $T$,
following the simple rule $\Gamma_T\approx 2\Gamma(\mu,T) \approx 2 T$,
if considered over the large range of $T=0\; -\; 6 \mev$). 
Two reasons may be responsible for this feature: (i) The
calculation still is performed within the (grand) canonical ensemble;
(ii) the choice of the self-energies in (\ref{dspspecdensgam}) does actually not
lift degeneracies.

It seems to me that the questions raised here suggest an interesting
and important problem for future studies, not just for our formulation
of transport theory but in the more general context. The problem at
stake touches two critical issues of nuclear physics: the use of both
the independent particle model as well as of the canonical ensemble. We
know that both models are not really applicable, and the evaluation of
static susceptibilities allows one to put the question on a
quantitative level. Indeed, might one not expect that a computation of
these on the basis of the compound model and for the micro-canonical
ensembles would show ergodicity, in the sense of (\ref{conderg})?

To demonstrate the importance of such questions for transport
properties let us look at nuclear friction. Often the latter may be
approximated by the so called zero-frequency limit, for which one has:
\bel{zefridisflu} \gamma(0) =
     {{\partial \chi^{\dpr} (\om)} \over{\partial \om}} 
      \Bigr\arrowvert_{\om = 0}=
     {\psi^{\dpr}(\om=0) \over 2T} =
  {2 \hbar \over \Gamma_T}\Bigl( \chi^{\rm T} - \chi(0) \Bigr)
   +{_R\gamma(0)}
\ee
In the expression on the very right, the first term represents
the contribution from the heat pole, the second one from the rest.
Estimating  $\chi^{\rm T} - \chi(0)$ through (\ref{heatzeroipm})
this component of friction turns into
the one found first by Ayick and N\"orenberg within the model of DDD
\cite{aynoediabfri} for larger temperatures. In Fig.5 it is shown by the
full and dashed lines  (see \cite{hoivyanp}) as function of $T$.
They correspond to values of the parameter $c$ of 
(\ref{dspspecdensgam}) put equal to $c=20$ MeV and $c\to \infty$,
respectively. The curve with the heavy squares corresponds to the
contribution ${_R\gamma(0)}$ of the remaining part of the correlation
function. The horizontal line represents wall friction.
As compared to experimental evidence from nuclear fission,
the friction force given by the (\ref{zefridisflu}) (with all terms
included) appears to be too big. Furthermore, one expects nuclear
friction to {\it increase} with excitation, rather than to decrease,
also in the regime of somewhat higher $T$.  From our discussion above
the culprit for such a misbehavior is easily traced back to a violation
of "ergodicity", in computations which base on the model of independent
particles in a (grand) canonical ensemble.  Therefore, it has been
argued in \cite{hoivyanp} and \cite{hofrep} to "re-install" the latter by
neglecting in (\ref{zefridisflu}) the contributions from $\chi^{\rm ad} -
\chi(0)$. For the "experimental facts" mentioned see e.g. \cite{higoroobn}
and \cite{hofbapa}, as well as the talk by G. Rudolf at this meeting in
which a new interpretation of such data has been presented. It is only
fair to add that at present no definite answer to such questions is
possible yet. However, there can be little doubt that they may be
considered as very important ones, not just for nuclear dissipation,
but to understand nuclear collective motion in the more general
context.

\begin{figure}[htb]
\centerline{
{\epsfysize=10cm 
\epsffile
{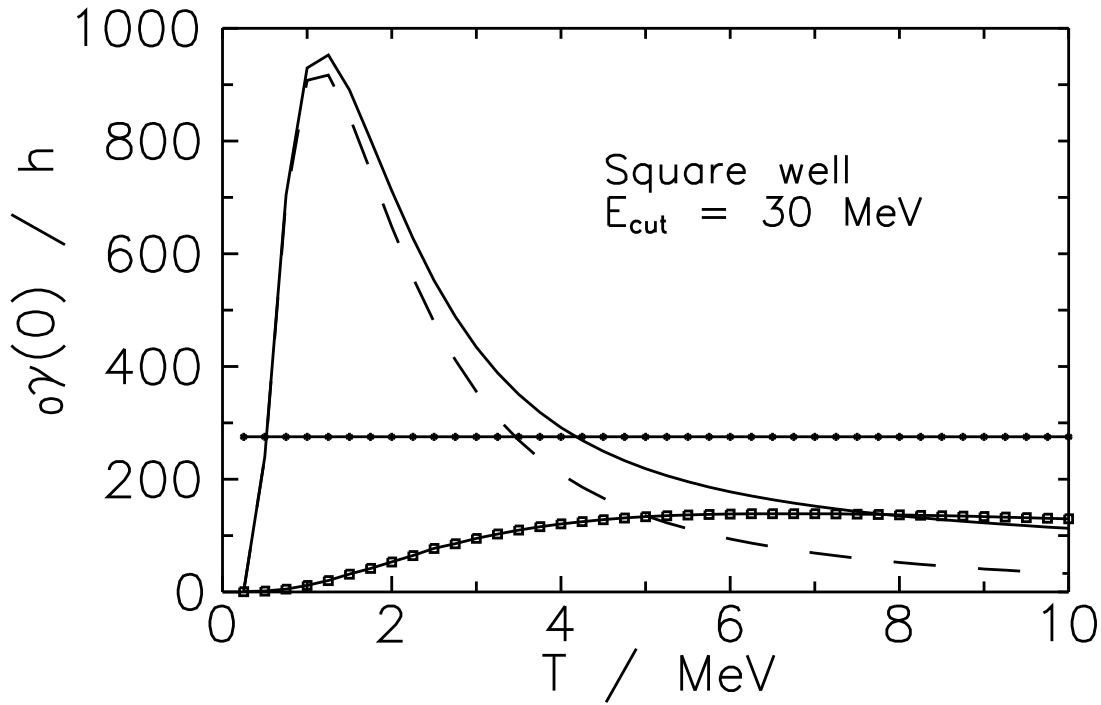}}}
\caption{\label{Fig.5}
Heat pole friction as function of temperature (see text)}
\end{figure}

\subsection{Fluctuation of transport coefficients with shape at small
excitations}

As seen from Fig.3, strong variations with $Q$ occur at the lower
temperatures of $1$ and $2\mev$. In principle, they show up because of
shell effects. Unfortunately, in practical applications they appear to
be grossly overestimated, simply because of unphysical irregularities
in the independent particle model. It seems that even pairing
correlations do not suffice \cite{ivhopair} to smooth out these
oscillations. This problem may hint to the need of using truly
{\it adiabatic states}. The latter would be produced after considering
in an improved way effects of the residual interaction $\Vres$. Apparently,
this feature is very much related to the one encountered before when
discussing susceptibilities. The $\Vres$ may be expected not only to lift
degeneracies but to repel states, not just at the spherical
configuration but for deformed states as well.

The puzzle raised here may be addressed from a different point of view:
It is very doubtful that such small scale variations may at all be
"seen" by the phase space distribution which moves
across the collective landscape. This $f(Q,P,t)$ itself will have some
natural width, both in $Q$ as well as in $P$. Therefore, one may well
question any use transport coefficients which exhibit
such extreme variations in the collective degrees 
of freedom. Obviously this problem is related to the
one raised by the late V.M. Strutinsky (see e.g. \cite{strutberlin}).
He questioned any introduction of collective degrees of freedom which
is not accompanied by a suitable averaging procedure. Please notice that the
variations we are talking about here are of a scale much smaller
than those one would see in the typical gross shell behavior. Indeed,
the ones shown in Fig.3 for $\gamma/M$ are related to those which in
the potential energy would occur in between the {\it few} maxima and
minima of a typical fission landscape (cf. e.g. \cite{ivhopaya}). As for
the transport coefficient of nuclear dissipation, the reader may
be reminded of the physics of the wall formula. The latter represents a
certain {\it macroscopic limit}, in which shell effects are washed out
totally (cf. \cite{hoivyanp} and \cite{hofrep}). This is the reason why no
dependence on $Q$ survives which might be related to shell effects of
any kind; this point is elaborated on in \cite{him}. 

\subsection{Breakdown of diffusion across a barrier at small
excitations}

It has been mentioned already at the end of sect.2 that diffusion
coefficients may be calculated even for unstable modes, which are
present at $C(0)<0$. This is by far not trivial as for such modes
the "equilibrium fluctuations" which appear in
(\ref{dkinqqpp})-(\ref{oscaverqqpp}) loose their physical meaning.
Such a calculation requires a delicate extension of linear response
theory and the FDT to instabilities; for an comprehensive discussion
see \cite{khqstsu}. Unfortunately the procedure breaks down at some
critical temperature $T_c$ below which the diffusion coefficient
$D_{pp}$ would become negative. This may be inferred from Fig.6,
together with (\ref{dkinqqpp}). \begin{figure}[htb]
\centerline{\rotate[r]{\epsfysize=12.5cm \epsfxsize=9.5cm 
\epsffile[95 73 549 671]{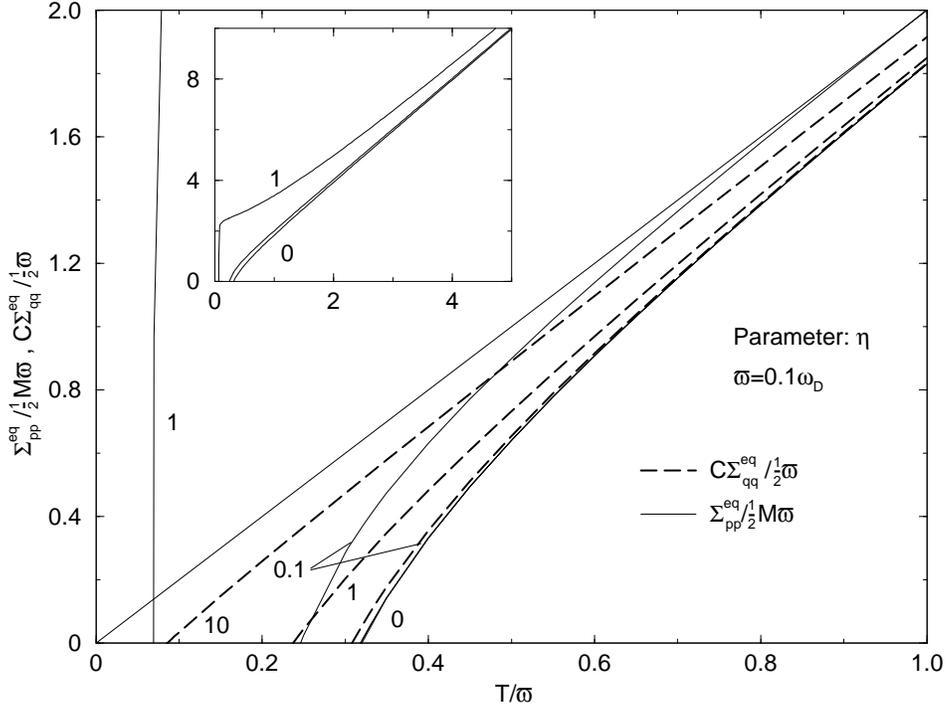}}}
\caption{\label{Fig.6}
Analytic continuations of the (diagonal) equilibrium
fluctuations to the case of negative stiffness (dashed line for the
coordinate, fully drawn lines for the momentum), as functions of
temperature for different values of friction. The insert shows the
momentum fluctuations for a larger range of the temperature and the
same values of $\eta$.
}
\end{figure}
The formal reason for this problem may be analyzed in the limit of weak
damping: For $C\to -C$ the $\coth$ in the $T^*$ introduced in
(\ref{efftemp}) simply turns into a $\cot$!  In this limit the $T_c$ is
larger than the so called "cross over" temperature $T_0$ found for
"dissipative tunneling" by  exactly a factor of 2. With increasing
$\eta$ both $T_0$ as well as $T_c$ become smaller (with their ratio
changing as well). For more details see \cite{khqstsu}.

As may be deduced from Fig.6 the problem occurs at temperatures of the
order of $T\approx 0.5
\mev$ or less. Thus it is in this regime where the concept of the LHA
may be said to fail, in regions where instabilities occur. However, as
demonstrated in \cite{hofthoming}, above this $T_c$ our method is capable
of accounting for quantum corrections to Kramers' decay rate of
fission, which without any doubt is a highly non-linear process.

\subsection{The problems with the entrance phase of a heavy ion
collision}

The theory discussed above was based on the assumption that it
is possible to devide the system into "fast" and "slow" degrees
of freedom. The former are supposed to exhibit genuine
relaxation to a "local" equilibrium on top of which motion of
the slow degrees of freedom may be described exploiting a
quasi static description. Such favourable conditions are hardly
fulfilled at the very early stages of heavy ion collisions, even
if the energy per particle does not exceed some few Mev. The
composite system starts from two separated fragments at zero
intrinsic excitation. It will take a finite time before a common
potential builts up with respect to which it may become
meaningful again to speak of thermal excitations. During this
phase the collective motion which is manifested in the distance
between the two center of masses of the two (more or less still
existing) fragments cannot be associated to low frequeny modes.
This feature has already been discussed in \cite{johan}. There
it has been demonstrated that a large part of the excitation may
be attributed to high frequency transitions in the nucleonic
response, which nevertheless may lead to a kind of
"pseudo-friction" for relative motion. In \cite{aynoediabfri}
non-Markovian equations of motion have been suggested to dwell
with this problem in the so called "diabatic dissipative
dynamics" (DDD). It is conceivable that such effects could
eventually be treated by generalizing the LHA suggested above in
such a way that one concentrates on high frequency
collective modes, rather than on the lowest one. After all, in
some schematic way, the local response calculated within the
DDD model looks very similar to strength distributions where
all more detailled excitations are neglected other than just one
prominant giant resonance, see \cite{hoivyanp}, \cite{kidhofiva},
\cite{hofrep}. A combination of this picture of
average motion with our way of handling the dynamics of
fluctuations might allow one to improve on the calculation of the
diffusion coefficients. This seems to be highly needed
as for the entrance phase the latter cannot be calculated within
the high temperature limit.

{\bf Acknowledgements:} The author greatfully acknowledges support from
the Deutsche Forschungsgemeinschaft. Furthermore, he likes to express
his deep gratitude to his colleagues F.A. Ivanyuk, D. Kiderlen, A.G.
Magner and S. Yamaji for their close collaboration over the past decade.

\vspace{2 cm}

\end{document}